\documentclass{article}
\usepackage{psfig}

\def\pderiv#1#2{
{{\partial #1} \over {\partial #2}}
}

\title{Pulse interaction in nonlinear vacuum electrodynamics}
\author{A. M. Ignatov\footnote{General Physics Institute, Moscow, Russia,
e-mail: aign@fpl.gpi.ru},
  V.P.Poponin
\footnote{International Space Sciences Organization, San
Francisco, CA, USA, \mbox{ e-mail: vpoponin@sj.znet.com  }}  }

\date{}
\begin{document}
\maketitle

\vspace{1 cm} \centerline{Abstract}

{\sl The energy-momentum conservation law is used to investigate
the interaction of pulses in the framework of nonlinear
electrodynamics with Lorentz-invariant constitutive relations. It
is shown that for the pulses of the arbitrary shape the
interaction results in phase shift only. }
\section{Introduction}

Although classical electromagnetic theory deals with linear Maxwell equations,
there have been numerous attempts to bring the nonlinear phenomena into the
stage. All relativistic and gauge invariant versions of electromagnetism are
based on the Lagrangian density, $L$, which depends on the invariants of the
field tensor. Generally, in terms of the electric ($\bf E$) and magnetic ($\bf
B$) fields the Maxwell equations in absence of external charges may be  written
in a standard  form:

\begin{eqnarray}
\bf{D}_t- \nabla \times \bf{H} =0,  \quad \nabla \bf{D}=0, \nonumber \\
\label{0}\\
\bf{B}_t+ \nabla \times \bf{E} =0,  \quad \nabla \bf{B}=0, \nonumber
\end{eqnarray}
where we put $c=1$ and $\bf{D}=\pderiv{L}{\bf{E}} ,\;
\bf{H}=-\pderiv{L}{\bf{B}} $.  The Lagrangian $L(I, J^2)$ depends on Poincare
invariants $I=\bf{E}^2-\bf{B}^2$ and $J={\bf E B}$ only. The distinctive feature
of Eqs.~(\ref{0}) is that
since the Poincare invariants are identically zero for the plane electromagnetic
wave, the latter is insensitive to vacuum nonlinearity and propagates without
distortion.

Of particular interest are the nonlinear corrections to the linear
electrodynamics arising due to vacuum polarization in the strong electromagnetic
field. In the ultimate case of slowly varying fields this results in Heisenberg-
Euler electrodynamics, which is discussed in many textbooks ({\sl e.g.}
\cite{akhiezer}).

The main point of this paper is to describe the simplest, in a sense, nonlinear
vacuum process: the interaction of two electromagnetic waveforms propagating in
opposite directions.

\section{Maxwell equations}

We consider a linearly polarized wave propagating in the $z$ direction of the
form $E_x=E(z,t), \; B_y=B(z,t)$ with all other components being zero. In this
situation, the second Poincare invariant vanishes, $J\equiv 0$, so the Maxwell
equations are written as

\begin{eqnarray}
\left( E L(I)_I \right)_t+\left( B L(I)_I \right)_z&=&0, \nonumber \\
  \label{meq} \\
\left( B \right)_t+\left( E\right)_z&=&0, \nonumber
\end{eqnarray}
where the subscript denotes the derivative with respect to the corresponding
variable and $I=E^2-B^2$. The Lagrangian in Eq.~(\ref{meq}) is expanded in
powers of $I$. Keeping the lowest-order nonlinear corrections we have

\begin{equation}
L(I)=I+\frac12 \sigma I^2 +\dots   \label{he}
\end{equation}
With the help of the appropriate scale transform, the coefficient
$\sigma$ may be reduced to $\pm 1$. For the particular case of the
Heisenberg-Euler electrodynamics, $\sigma=1$ \cite{akhiezer}. Of
interest also is to keep in mind the Born-Infeld electrodynamics
({\sl e.g. } \cite{tonnela}) with the Lagrangian
\begin{equation}
L_{BI}(I)= 1- \sqrt{1-I}.  \label{bi}
\end{equation}

\section{Energy-momentum tensor}

The conservation laws for Eqs.~(\ref{meq}) are given by
\begin{equation}
W_t+N_z=0, \qquad N_t+P_z=0, \label{cons}
\end{equation}
where the components of the energy-momentum tensor, namely, the energy density,
$W$, the momentum density, $N$, and the stress, $P$, may be obtained using
standard variation procedure ({\it e.g.} \cite{landau}). Explicitely,

\begin{eqnarray}
W&=&2 E^2 L_I-L \nonumber \\
N&=&2 EB L_I  \label{emt}\\
P&=&2 B^2 L_I+L. \nonumber
\end{eqnarray}

Usually Eqs.~(\ref{cons},\ref{emt}) are thought of as a
 consequence of the Maxwell equations (\ref{meq}).
However, we may consider the relations (\ref{emt})
as a constraint implied upon the components of the momentum-energy tensor,
so there are two independent variables in Eqs.~(\ref{cons}), for example,
$W$ and $N$. One can easily check that for the nontrivial solutions of
Eqs.~(\ref{meq}), {\sl i.e.} for $I\neq 0$, the Jacobian of the transform
$E,B \to W,N$ is non-zero. Thus, instead of looking for the solutions of
 Eqs.~(\ref{meq}) we can solve Eqs.~(\ref{cons},\ref{emt}) excluding
the Poincare invariant $I$ from Eqs.~(\ref{emt}).

\section{Solution}

To exclude $I$ it is convenient to introduce the invariants of the energy-
momentum tensor, that is, its trace, $S=P-W$, and the determinant $T=WP-N^2$. As
it follows from Eqs.~(\ref{emt})
\begin{eqnarray}
S&=&2(L-I L_I) \nonumber\\ \label{inv} \\ T&=&I \left(L^2
\right)_I-L^2. \nonumber
\end{eqnarray}
The latter relations implicitly define the dependence $T=T(S)$. Substituting the
Lagrangian (\ref{he}) into Eqs.~(\ref{inv}) we find that the first nonvanishing
term of the expansion of $T$ in powers of $S$ is linear and it is provided by
the quadratic term of the expansion (\ref{he}): $T(S)=-\sigma S +\dots$. It is
noteworthy that the Born-Infeld Lagrangian (\ref{bi}) yields exactly the linear
dependence $T(S)=-S$.

The relations (\ref{cons}) are resolved introducing the potential $\psi$:
$W=\psi_{zz},\; N=-\psi_{zt},\; P=\psi_{tt}$. Restricting  ourself with the
linear relation between $T$ and $S$, we obtain the Ampere-Monge type equation
for $\psi$:
\begin{equation}
\psi_{zz}\psi_{tt}-\psi_{zt}^2 =\sigma(\psi_{zz}-\psi_{tt}).  \label{am}
\end{equation}

There are trivial solutions to this equation $\psi(z,t)=F(z\pm t)$ with an
arbitrary function $F$, which correspond to the plane electromagnetic waveforms
described by Eqs.~(\ref{meq}) with $I=0$.
 Besides these, implementing the Legendre transform \cite{courant}
one can easily obtain the general integral of Eq.~(\ref{am}) valid
for $T\neq 0$ and, consequently, for $I\neq 0$.
As a result, we get the components of the energy-momentum tensor
in a parametric form:

\begin{eqnarray}
W&=&\sigma(F_1^\prime(\xi)+ F_2^\prime(\eta)+2 F_1^\prime(\xi)
F_2^\prime(\eta))/\Delta(\xi,\eta), \nonumber \\
 P&=&\sigma(F_1^\prime(\xi)+ F_2^\prime(\eta)-2 F_1^\prime(\xi)
F_2^\prime(\eta))/\Delta(\xi,\eta), \nonumber \\
N&=&\sigma(F_2^\prime(\eta)- F_1^\prime(\xi) )/\Delta(\xi,\eta),
\label{sol} \\
z&=&\frac12 (\xi+\eta-F_1(\xi)-F_2(\eta)), \nonumber \\
t&=&\frac12 (\xi-\eta+F_1(\xi)-F_2(\eta)), \nonumber
\end{eqnarray}
where $F_{1,2}$ are arbitrary functions and $\Delta(\xi,\eta)=1- F_1^\prime(\xi)
F_2^\prime(\eta)$.

Consider, for example, two localized pulses of the arbitrary shape propagating
in opposite directions. This corresponds to the following initial conditions:

\begin{equation}
W(z,t)|_{t\to -\infty} = W_1(z+t)+W_2(z-t),  \label{ini}
\end{equation}
where $W_{1,2}(\xi)|_{\xi \to \pm\infty} \to 0$.
This initial condition is provided by the following choice of
$F_{1,2}$ in Eqs.~(\ref{sol}):$ F_{1,2}^\prime(\xi)=\sigma W_{1,2}(\xi)$ and
$F_1(\xi)|_{\xi\to-\infty}\to0$,
$F_2(\xi)|_{\xi\to\infty}\to0$. The asymptotic of the solution (\ref{sol}) at
$t\to\infty$ is then given by

\begin{equation}
W(z,t)= W_1(z+t-\sigma K_1)+W_2(z-t+\sigma K_1), \label{as}
\end{equation}
where
\begin{equation}
K_{1,2}=\int\limits_{-\infty}^\infty d\xi\; W_{1,2}(\xi) \nonumber
\end{equation}
is the net energy carried by the corresponding pulse.

Typical plots $W(z,t)$ are depicted at Fig.~1,2. The first picture shows the
interaction of two identical pulses. The interaction of one pulse with a
sequence of two pulses is shown in Fig.~2.

\section{Discussion}

Of interest is the geometrical sense of the obtained solution (\ref{sol}). The
parameters $\xi$ and $\eta$ are, in fact, the light-cone coordinates disturbed
by the electromagnetic field. One may say that the electromagnetic field alters
the space-time metric due to the dependence of the speed of light on the field
strength. In contrast with general relativity, the space-time remains flat.

Another interesting point is that for $\sigma=1$ the increase in the pulse
amplitude results in delay in energy (and information) exchange between distant
points, that is, the solution described by (\ref{as}) is subluminal.
This takes place for both the Heisenberg-Euler electrodynamics, which is
currently the only one of physical sence,  and  for the elegant Born-Infeld
theory, for which our results are exact. However, for $\sigma=-1$ the pulse
propagation would be superluminal.

From the viewpoint of nonlinear physics, the electromagnetic pulses  in vacuum
exhibit the soliton-like behavior: the collision results in a phase shift but
the form of a pulse remains unchanged. The main interesting point with this
respect is that unlike usual nonlinear equations, the shape of the soliton is
arbitrary.

\newpage
\begin{figure}
\begin{center}
\psfig{figure=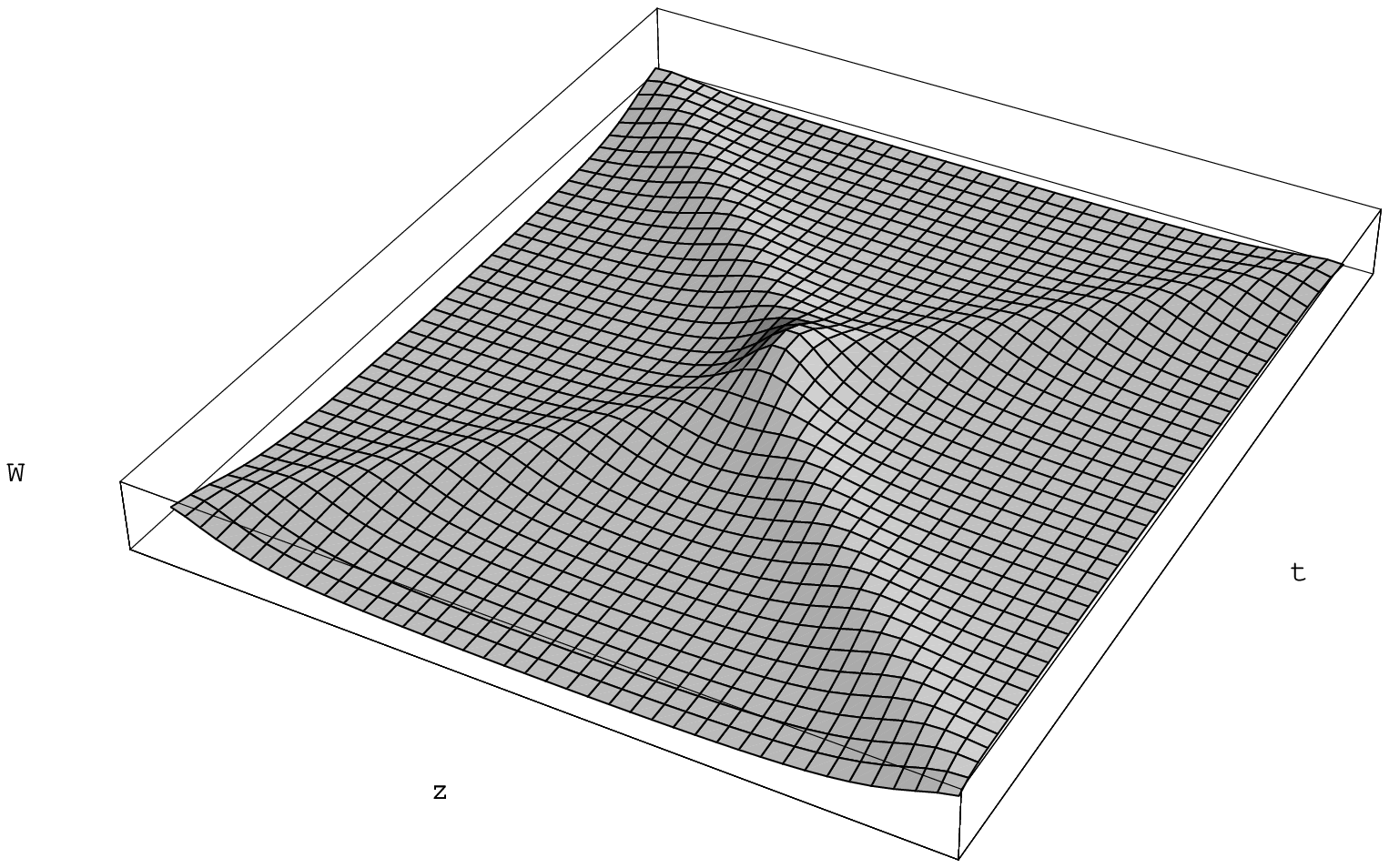,width=10cm} \caption{Interaction of two
identical pulses}
\end{center}
\end{figure}

\newpage
\begin{figure}
\begin{center}
\psfig{figure=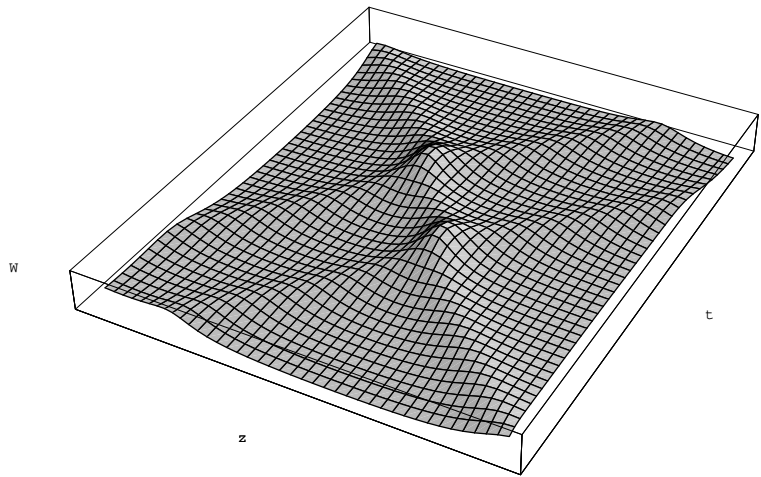,width=10cm} \caption{A single pulse
interacting with a sequence of two pulses}
\end{center}
\end{figure}

\end{document}